\documentclass[12pt]{article}

\usepackage[letterpaper,hmargin=1in,vmargin=1in]{geometry}

\usepackage{graphicx,amsmath,amsfonts}

\def\be{\begin{equation}}
\def\ee{\end{equation}}
\def\ba{\begin{eqnarray}}
\def\ea{\end{eqnarray}}
\def\ge{\mathrel{\raise.3ex\hbox{$>$\kern-.75em\lower1ex\hbox{$\sim$}}}}
\def\la{\mathrel{\raise.3ex\hbox{$<$\kern-.75em\lower1ex\hbox{$\sim$}}}}

\def\theequation{\thesection.\arabic{equation}}
\def\simgt{\mathrel{\raise.3ex\hbox{$>$\kern-.75em\lower1ex\hbox{$\sim$}}}}
\def\simlt{\mathrel{\raise.3ex\hbox{$<$\kern-.75em\lower1ex\hbox{$\sim$}}}}

\newcommand{\bi}[1]{\bibitem{#1}}

\newcommand{\R}{\mbox{$\fr{1-\gamma_5}{2}$}}

\newcommand{\nc}{\newcommand}

\nc{\gone}{\bar g_{\pi NN}^{(1)}}
\nc{\gzero}{\bar g_{\pi NN}^{(0)}}
\nc{\al}{\alpha}
\nc{\ga}{\gamma}
\nc{\de}{\delta}
\nc{\ep}{\epsilon}
\nc{\ze}{\zeta}
\nc{\et}{\eta}

\nc{\Th}{\Theta}
\nc{\ka}{\kappa}
\nc{\rh}{\rho}
\nc{\si}{\sigma}
\nc{\ta}{\tau}
\nc{\up}{\upsilon}
\nc{\ph}{\phi}
\nc{\ch}{\chi}
\nc{\ps}{\psi}
\nc{\om}{\omega}
\nc{\Ga}{\Gamma}
\nc{\De}{\Delta}
\nc{\La}{\Lambda}
\nc{\Si}{\Sigma}
\nc{\Up}{\Upsilon}
\nc{\Ph}{\Phi}
\nc{\Ps}{\Psi}
\nc{\Om}{\Omega}
\nc{\ptl}{\partial}
\nc{\del}{\nabla}
\nc{\ov}{\overline}
\nc{\newcaption}[1]{\centerline{\parbox{15cm}{\caption{#1}}}}
\nc{\none}{${\cal N}=1\;$}
\nc{\ntwo}{${\cal N}=2\;$}
\nc{\nfour}{${\cal N}=4\;$}
\nc{\nones}{${\cal N}=1^*\;$}
\nc{\Z}{\mathbb{Z}}
\renewcommand{\R}{\mathbb{R}}

\begin{document}

\begin{titlepage}

\rightline{hep-ph/0606050}

\setcounter{page}{1}

\vspace*{0.2in}

\begin{center}

\hspace*{-0.6cm}

\Large{\bf Central charges, S-duality and massive vacua of \nones super Yang-Mills}

\vspace*{0.5cm}
\normalsize

{\bf Adam Ritz}

\smallskip
\medskip

{\it Department of Physics and Astronomy, University of Victoria, \\
     Victoria, BC, V8P 5C2 Canada}

\smallskip
\end{center}
\vskip0.4in

\centerline{\large\bf Abstract}

We provide a simple derivation of the extremal values of the superpotential in massive vacua of \nones SYM, making use of the
required modular weight for the central charge of BPS walls interpolating between these vacua. 
This modular weight  descends from the action of S-duality on the \nfour superalgebra which in turn is inherited from its classical action on 
the dyon spectrum. We show that this kinematic information, combined with 
minimal knowledge of the  weak coupling asymptotics, is sufficient to
determine the exact vacuum superpotentials in terms of Eisenstein series.

\vfil
\leftline{June 2006}

\end{titlepage}

\renewcommand{\theequation}{\arabic{equation}}

Maximally supersymmetric \nfour super Yang-Mills exhibits many remarkable features, among which its invariance under
electric-magnetic duality is one of the most profound \cite{mo}. In the conformal phase this symmetry acts on the dimensionless coupling
constant of the theory,
\be
 \ta \longrightarrow \frac{a\ta+b}{c\ta +d}, \;\;\;\;\; {\rm with}\; a,b,c,d \in \Z,\; ad-bc=1,
\ee
and thus the coupling can be taken to lie in the fundamental domain of SL$(2,\Z)$. If one moves away from the conformal
point onto the Coulomb branch, this symmetry acts as a generator of the spectrum of massive BPS states. More
precisely, on branches of the moduli space where only one of the three adjoint scalar fields $\ph^a$ has a nonzero vev,
the BPS spectrum \cite{wo},
\be
 M = |{\cal Z}|,\;\;\; {\cal Z}= \sqrt{\frac{2}{{\rm Im}\ta}}(n_e^a + \ta n_m^a)\ph_a,
\ee
is permuted by the action of SL(2$,\Z$), where
\be
 \left(\begin{array}{c} n_e \\ n_m \end{array}\right) \longrightarrow 
  \left(\begin{array}{cc} a & -b \\  -c & d \end{array}\right)
  \left(\begin{array}{c} n_e \\ n_m \end{array}\right). 
 \ee
 
 This action of $S$-duality on the BPS spectrum is of course well-known. However, it has some interesting and less well-explored consequences. 
 In particular, as shown by Intriligator \cite{intril} and recently discussed by 
 Kapustin and Witten \cite{kw}, there is an induced action on the supersymmetry algebra. For our purposes, it will be convenient to 
 focus on  the anticommutator of two left-handed 
 supercharges in the ${\cal N}$--extended algebra, which contains two sets of allowed central charges,
 \be
  \{ Q^A_\al, Q^B_\beta\} = \ep_{\al\beta} Z^{[AB]} + \si^{\mu\nu}_{\al\beta} Z^{(AB)}_{\mu\nu}.
 \ee
 The first charge here appears only for ${\cal N}\geq 2$, and relates to the dyonic BPS spectrum considered above \cite{wo}. The second
 charge is supported by BPS domain walls \cite{fmvw,at,ds} which, while not present in \nfour SYM, do
 arise in many \none theories to which \nfour SYM flows under relevant perturbations.
 
 In this note we will assume the exact invariance of \nfour SYM under SL(2,$\Z$)  and explore the ensuing consequences for the
 tensorial central charges ${\cal Z}_{\mu\nu}$ in the superalgebra. In particular, the required covariance of
 ${\cal Z}_{\mu\nu}$ under modular transformations imposes stringent constraints on the low energy superpotentials in massive vacua 
 which result from  \none perturbations of \nfour$\!$. In particular, for the \nones deformation, we will be able to compute the exact vacuum superpotentials simply by requiring the necessary modular properties, given  some cursory knowledge of the weak coupling 
 asymptotics.  
 
 The primary constraint we require follows straightforwardly from the transformation rules above. In particular, we 
 observe that the Lorentz-scalar central charge ${\cal Z}$ transforms with modular
 weight,
 \be
  w({\cal Z}) = (-1/2,1/2),
 \ee
where the notation, $w(f)=(w_1,w_2)$, implies  $f((a\ta+b)/(c\ta+d)) = (c\ta+d)^{w_1} (c\bar{\ta}+d)^{w_2} f(\ta)$.
 It is then apparent from the structure of the superalgebra that if we deform the theory with a relevant perturbation, breaking \nfour to \none SUSY and leading to massive vacua and thus the possibility for BPS domain walls, the modular weight of the central charge
${\cal Z}_{\mu\nu}$ must again be 
\be
 w({\cal Z}_{\mu\nu}) = (-1/2,1/2). \label{wt}
\ee
This can be seen from the fact that the modular weight of the unbroken supercharges is already fixed from their embedding within the
\nfour algebra above. 

Eq.~(\ref{wt}) is the primary result that we will exploit in the remainder of this note. In particular, this central charge
can generically be expressed in  the form,
\be
 {\cal Z} = \De\!\!\left.{\cal W}\right|_v,
\ee
with $\left.{\cal W}\right|_v$ the extremal value of the low energy superpotential in each vacuum between which the wall 
interpolates. Thus, we can conclude that the extrema of the superpotential (possibly corrected by a vacuum-independent constant) also inherit the same modular weight,
\be
 w(\left.{\cal W}\right|_v) = (-1/2,1/2). \label{modc}
\ee
The possibility for a vacuum-independent constant to be added so that ${\cal W}$ and not just $\De{\cal W}$ is
modular can be associated with operator mixing \cite{adk}.

If we now choose a particular relevant deformation of the theory, this result constitutes 
a powerful constraint on the extremal values of the superpotential. To proceed, we consider the \nones
deformation, for which the classical  superpotential takes the form,
\be
 {\cal W} = N{\rm Tr}\left[ \Ph_1 [\Ph_2,\Ph_3] + \frac{1}{2}m \sum_{i=1}^3 \Ph_i^2\right].
\ee
The exact extremal values for the superpotential in the massive vacua of this system were first obtained explicitly by Dorey \cite{dorey} (see also
\cite{dw,dk,ps,adk,dhks,klyy}) and are well-known, but we will 
illustrate this technique by rederiving these results in a very straightforward manner which will also serve to illustrate
the extent to which these results are determined purely by kinematics. Indeed, we will limit the dynamical input to knowledge of the
classical Higgs vacuum, which is given by expressing $\Ph_i$ in terms of the unique irredicible representation of 
SU(2) of dimension $N$ (taking the gauge group to be SU($N$)). In the normalization above, we have
\be
 \left.{\cal W}\right|^{\rm cl}_{h} = \frac{N^3}{24}m^3\left[ (-N) + {\cal O}(1) + {\cal O}(e^{2\pi i N\ta})\right], \label{higgscl}
\ee
which is valid in the weak coupling regime, $\ta\rightarrow i\infty$. The ${\cal O}(N^3)$ prefactor results 
directly from our normalization of the superpotential. We have retained only the leading constant term for large $N$, as this allows us to
exclude possible operator mixing ambiguities which vanish at large $N$. We have also exhibited the scaling of the leading nonperturbative correction,
an $N$-instanton contribution. The origin of this scaling follows first of all from the fact that, since this vacuum can be reliably placed at weak coupling,
we expect the nonperturbative contributions to be exhausted by instantons. Secondly, the absence of  $k$-instanton contributions for $k<N$
is most clearly understood from an analysis of the ADHM constraints, or more directly from the
realization of the relevant instanton configurations in terms of a D(-1)-D3 system. For instantons to contribute to the superpotential, the 
worldvolume theory of the D(-1)-branes must have a supersymmetric vacuum and, as shown in \cite{kh}, in the presence of the 
\nones deformation the $F$-term  constraints are only satisfied if the number of instantons is a multiple of $N$.  We are thus led to the
above scaling of semi-classical contributions in the Higgs vacuum. These results will be of use below.

To proceed in making use of the general constraint (\ref{modc}), we need to determine the modular weight of the adjoint mass 
parameter $m$. This can be done by returning to the conformal phase and requiring that the chiral primaries be modular 
invariant, from which it follows that \cite{adk}
\be
 w(m) = (-5/6,1/6). \label{wtm}
\ee

Since the effective superpotential in each vacuum must vanish as the deformation $m\rightarrow 0$, we can
use global symmetries and dimensional analysis to write
\be
 \left.{\cal W}\right|_v = \frac{N^3}{24} m^3 X(\ta)|_v,
\ee
with an unknown function $X(\ta)$, depending only on the (bare) coupling $\ta$. By holomorphy of the superpotential it 
follows that $X$ must be a holomorphic modular form, and indeed using (\ref{wtm}) we see that 
\be
 w(X) = (2,0),
\ee
for consistency with the modular weight of the central charge.

This leads us to conclude that the nontrivial $\ta$-dependence of the vacuum condensates in any massive vacuum of \nones SYM must be 
determined by a suitable holomorphic weight-2 modular form of SL(2,$\Z$). Unfortunately, this neat conclusion cannot be
correct as there are no such forms. There is a unique candidate which comes closest, namely the regulated second Eisenstein
series,
\be
 E_2(\ta) \equiv \frac{3}{\pi^2} \sum_{(a,b)\,\in\, \Z^2-\{0,0\}}\frac{1}{(a\ta + b)^2},
\ee 
which however transforms under $\ta \rightarrow -1/\ta$ as weight-2 only up to an additive shift.

In fact, this conclusion shouldn't be a surprise as the dyonic central charge ${\cal Z}(\ta)$ is 
in fact only a modular form of weight $(-1/2,1/2)$ up to {\it permutation}, as a suitable action on the electric and magnetic 
charges, $(n_e,n_m)$, was also required. Therefore, we should anticipate a similar structure in the present case. Indeed, physically we expect these
massive vacua to be associated with the condensation of various dyonic states, and thus should directly inherit this permutation
under SL(2,$\Z$). 

It follows that a given massive vacuum will only preserve a specific subgroup $\Ga\subset\,$SL(2,$\Z$), and the broken generators will induce the
required permutation. We could proceed by trying to determine the precise subgroup $\Ga$; the required forms will be of weight-2
with respect to this subgroup. However, it is clear that a priori there is no reason to believe that all massive vacua preserve the same
subgroup, or equivalently that all vacua lie on the same orbit of the associated coset. Instead we will construct the full orbit explicitly
without making an assumption about the residual subgroup in any given vacuum. This clearly requires some dynamical information,
but it will be sufficient to use knowledge of the weak coupling asymptotics of the Higgs vacuum discussed above.

To see how this works, recall that weight-2 modular forms for general (congruence) subgroups of SL(2,$\Z$)  
can be constructed in terms of the basis of forms for the principal congreunce subgroups (see e.g. \cite{koblitz}), 
\be
 \Ga(M) \equiv \left\{ \left(\begin{array}{cc} a & b \\ c & d \end{array}\right) \in {\rm SL(2,}\Z), \;\; a=d=1\; ({\rm mod}\; M), \; b=c=0\; ({\rm mod}\; M) \right\}.
 \ee
 which (for weight-2) are suitable linear combinations of  `level $M$' Eisenstein series
\be
 G_2^{\vec{v}}(\ta) \equiv \sum_{(a,b) =(\vec{v}\; {\rm mod}\; M)\, \in \Z^2 - \{0,0\}} \frac{1}{(a\ta + b)^2},
 \ee
 which make use of a vector $\vec{v}=(v_1,v_2)$ with $1\leq v_i\leq M$. 
  
 The vectors $\vec{v}$ are in one-to-one correspondence with the cusps of $\Ga\subset\,$SL(2,$\Z$). Note that although 
 $G_2(\ta)$ is not a strict modular form itself, the additive  shift under $\ta \rightarrow -1/\ta$ can be cancelled by taking a suitable linear combination,
 given a  subgroup $\Ga$ with at least two cusps. 
  
 For a fixed vector $\vec{v}$, generic  elements of SL(2,$\Z$) will permute the basis forms via the natural action
 \be
  \vec{v} \rightarrow \{ \vec{v} \ga | \ga \in {\rm SL(2,}\Z)\}.
 \ee
 To pin down this action precisely, we need to find one point on each orbit. As noted above, it will be sufficient to use the Higgs vacuum
 which is visible at weak coupling. Expanding the Eisenstein series in the weak coupling $\ta \rightarrow i\infty$ limit, we find 
 \be
  G_2^{\vec{v}} \longrightarrow c_1\de_{v_1 0}+\frac{c_2}{M^2}e^{2\pi i (\ta n+v_2 m)/M} + \cdots,
 \ee
 where $c_1$ and $c_2$ are constants that will not be important in what follows. The leading nonperturbative correction shown here is determined
 by the constraint on the integers $n\geq 1$ and $m | n$, namely that $n/m \equiv v_1$ mod $M$. Matching the asymptotics of the Higgs 
 vacuum, $X(\ta)|_h \rightarrow (-N) + {\cal O}(1) + {\cal O}(e^{2\pi i N \ta})$, requires $\vec{v} = (0,v_2)$ in 
 order to retain the constant term. In this case, $n/m \equiv 0$ mod $M$, and thus to ensure that
 the leading nonperturbative correction is no larger than the $N$-instanton factor, we need to set $M=N$, i.e. given by the rank of the gauge group, 
 and sum over all the allowed values of $v_2$. Up to this point $M$ was simply a parameter 
 labelling the allowed set of modular forms, but with hindsight a relation to the rank of the gauge group is perfectly natural in 
 the sense that fully Higgsed vacua, which are visible at weak coupling, are unique for each SU($N$) gauge group. 
 
 We are thus led to a unique possibility:
 \be
  \frac{3}{\pi^2} \sum_{v_2=0}^{N-1} G_2^{(0,v_2)} = E_2(N\ta). \label{higgs}
 \ee
This combination is not modular, but a simple linear combination which cancels the additive shift under $\ta \rightarrow -1/\ta$ is given by
\be
 E_2(N\ta) - \frac{1}{N} E_2(\ta).
\ee 
To match the constant asymptotic value at large $N$, we can fix the prefactor and identify,
\be
 \left.X(\ta)\right|_h = E_2(\ta) - NE_2(N\ta),
\ee
which indeed coincides with the exact expression for the superpotential in the Higgs vacuum, determined previously 
using other approaches \cite{dorey,dk,dhks}.
  
 Alternatively, this result could have been deduced by `guessing' that the Higgs vacuum, since it is unique, would preserve the 
 largest congruence subgroup. Identifying
 the level with the rank $N$ as above, these subgroups are given by,
\be
 \Ga_0(N) \equiv \left\{ \left(\begin{array}{cc} a & b \\ c & d \end{array}\right) \in {\rm SL(2,}\Z), \;\; c = 0\; ({\rm mod}\; N)\right\}.
\ee
For a given prime integer $N$, these subgroups have precisely two cusps and therefore a {\it unique} modular form of weight 2, which 
is conventionally expressed in terms of the second Eisenstein series in precisely the combination deduced above \cite{koblitz},
 \be
  E_{2,N}(\ta) = N E_2(N\ta) - E_2(\ta).
 \ee
 
 We can now determine the remaining massive vacua from the orbit of the Higgs vacuum under SL(2,$\Z$). The sum in (\ref{higgs}) ensures that
 the subgroup $\Ga$ preserved by the vacuum is of index $N$ within SL(2,$\Z$), and we will henceforth assume that all massive vacua lie on the
 orbit of the Higgs vacuum. The action of SL(2,$\Z$) will preserve this index, and thus it is straightforward to write down
 the remaining possibilities, which are given by summing over $v_1$ and $v_2$ with fixed index $N$. i.e., for $N=pq$,
 \be
  \frac{3}{\pi^2} \sum_{v_1=0}^{q-1} \sum_{v_2=0}^{p-1} G_2^{(pv_1,qv_2)} = \frac{1}{q^2} E_2\left(\frac{p\ta}{q}\right).
 \ee
 Taking the same linear combination to restore the modular transformation properties, we obtain the candidate vacua,
 \be
  \left. X(\ta)\right|_{p,q} = E_2(\ta) - \frac{p}{q} E_2\left(\frac{p\ta}{q}\right),
 \ee
 which (with the shifts $\ta \rightarrow \ta+k$, $k=0..q-1$, induced by $2\pi$-rotations of the UV $\theta$ parameter) reproduce all the 
 known massive vacua of \nones SYM \cite{dw,dorey,dk,ps,adk}. For $N$ prime, the choice $(p,q)=(N,1)$ 
 reproduces the Higgs vacuum discussed above,
 while the alternative $(p,q)=(1,N)$ has the appropriate scaling to be identified with gaugino codensation in confining vacua.
 In general,  the apearance of the vector $\vec{v}$ is natural here in the context of 't Hooft's $\Z_N \times \Z_N$ classification of massive 
 phases \cite{thooft,thooft2}, 
 as it inherits a natural action of this group from the remaining group elements of SL(2,$\Z$) which thus permute all the massive phases.
  
 One should bear in mind that the procedure we have followed can in principle only determine the relative differences 
 between superpotentials in different vacua. However, in the present case, this  ambiguity has been fixed by the additional 
 assumption of modular covariance of ${\cal W}$, rather than just $\De {\cal W}$. 
 
 We will finish with some additional remarks on these results.
 
 \begin{itemize}
 \item The orbit of the Higgs vacuum under the broken generators of SL(2,$\Z$) has an interesting interpretation in terms of the Hecke 
 $T_N$ operators, which map the space of weight-$k$ forms into itself by a suitable averaging procedure. More precisely, 
 if $f$ is a modular form of SL(2,$\Z$), the action of the Hecke operator $T_N$ is given by (see e.g. \cite{koblitz})
\be
 T_N f(\ta) = \sum_{pq=N, k=1..N} \frac{p}{q}f \left(\frac{p\ta+k}{q}\right),
\ee
with the sum over the divisors $p$ of $N$ and $k=1..N$.
The forms that are relevant here are eigenvectors of $T_N$ of weight-2, of which there is only one, the regulated second Eisenstein series,
and
\be
 T_N E_2(\ta) = \si_1(N) E_2(\ta),
\ee
where $\si_1(N) = \sum_{d | N} d$ sums the divisors of $N$. Thus we can identify the orbit of the Higgs vacuum under SL(2,$\Z$) with the
orbit generated by $T_N$. Indeed, this correspondence is less mysterious given that both averages must restore full SL(2,$\Z$) covariance.
 \item In recent work, Kapustin and Witten \cite{kw} pointed out a relation between specific geometric Hecke operators and the insertion of 
 't Hooft operators. Since the latter also provide a shift of the vacuum analogous to the action of SL(2,$\Z$), it would interesting to understand
 the relation to this aspect of \cite{kw} in more detail.
 \item There is an alternative definition of the action of the Hecke operator $T_N$ using lattices. In particular, if the lattice associated with SL(2,$\Z$)
 has periods $(\om_1,\om_2)$ with $\ta=\om_2/\om_1$, then the Hecke operator acts by summing over all sublattices of index $N$. This is precisely
 the picture of massive vacua in \nones that 
 emerges from perturbations of \ntwo SYM \cite{dw} and also compactification on $\R^3\times S^1$ \cite{dorey}.
 \end{itemize}

 In conclusion, the approach we have outlined is quite general, and it would be interesting to know whether it can usefully be applied to deduce the
 vacuum superpotentials for other relevant perturbations of \nfour SYM, or indeed perturbations of other conformal theories, such as
 \ntwo SQCD with $N_f=2N_c$, $\beta$-deformations of \nfour, and other examples which are believed to admit an action of SL(2,$\Z$).

\subsection*{Acknowledgements}
I would like to thank N. Dorey for helpful comments on the manuscript. This work was supported in part by NSERC, Canada.

\end{document}